\documentclass[preprint,showpacs,preprintnumbers,amsmath,amssymb]{revtex4}
\usepackage{color}

\usepackage{graphicx}
\usepackage{dcolumn}
\usepackage{bm}
\usepackage{amssymb}
\graphicspath{{plots/}}

\begin{document}

\title{Reply to comment of Shukla et al. on \\`On ``Novel attractive forces'' between ions in quantum plasmas -- \\ failure of linearized quantum hydrodynamics'}

\author{M. Bonitz, E. Pehlke, and T. Schoof}%
\affiliation{%
    Christian-Albrechts-Universit\"at zu Kiel, 
    Institut f\"ur Theoretische Physik und Astrophysik, 
    Leibnizstra\ss{}e 15, 24098 Kiel, Germany
}

\date{\today}

\begin{abstract}
\end{abstract}

\maketitle

%\section{Introduction}\label{s:intro}

In our original paper, Ref.~\cite{bonitz_comment}, we tested the predictions of Shukla et al. in 
Ref.~\cite{shukla_11} for dense hydrogen, and we came to the following two main conclusions: 
(1) the prediction of a ``novel attractive proton-proton potential'' is wrong and (2), this error arises 
from a failure of the linearized quantum hydrodynamics (LQHD) model applied by the authors 
outside its applicability range. This conclusion was based on a comparison with our 
density functional (DFT) simulation results.
In their comment \cite{shukla_comment}, Shukla, Eliasson, and Akbari-Moghanjoughi (SEA), in fact, have 
come to very similar conclusions about the validity of the LQHD. At the same time there appears 
to be a misunderstanding about the physical basis of DFT that apparantly has led SEA to misinterpret 
our results. For the benefit of the reader we, therefore, give a brief comparison of the two 
methods below, critically assessing their respective strengths and weaknesses.

%\widetext

\begin{center}
\begin{table}
\begin{tabular}{|c|l|c|c|}
\hline
  & {\bf Property} & {\bf DFT} & {\bf QHD, LQHD}  
\\\hline
(1) & Quantum diffraction effects & included for & approximate, via Bohm potential\\
&&Kohn-Sham states & (``quantum recoil'') \cite{shukla_comment} \\
(2) & Quantum coherence effects & included for & missing \\
  & {\it e.g.} quantum interference & Kohn Sham states & (no phase information) \\
\hline
(3) & Spin effects, Pauli principle & approximate & approximate, via ideal equation of state \\
&&via $E_{\rm XC}$ & (``Fermi pressure'') \cite{shukla_comment} \\  
(4) & Many-particle effects   & approximate  & missing (standard formulation)\\
&& via $E_{\rm XC}$& corrections derived from $E_{\rm XC}$ \cite{manfredi08,shukla_11}$^{\rm (a)}$\\
(5) & Accessible temperature & $T=0$ $^{\rm (b)}$ & $T=0$\\
(6) & Accessible density     & no restriction  & weak coupling, $\quad r_s \lesssim 1$\\
&&& incomplete: misses Friedel oscillations \\
\hline
(7)  & Resolvable length scales & no restriction & $l > {\rm several}\quad {\bar r}$\\
(8)  & Strength of perturbation & no restriction & weak, $|\delta n| < n_0$\\
\hline
(9)  & Computational effort  & large  & low, semi-analytical \\
\hline
\end{tabular}
\caption{Comparison of key properties and limitations of standard density functional theory 
(DFT) and linearized quantum hydrodynamics (LQHD).
$E_{\rm XC}$: exchange-correlation functional, $r_s = {\bar r}/a_{\rm B}$, is the quantum coupling 
(Brueckner) parameter,
where ${\bar r}$ denotes the mean interparticle distance and $a_{\rm B}$ the Bohr radius. 
$n_0$ is the unperturbed density and $\delta n$ the induced density response. Comments: $^{\rm (a)}$ these references used a simplified version of $E_{\rm XC}$ from DFT. $^{\rm (b)}$ Extensions to finite temperature exist. 
}
\label{tab:1}
\end{table}
\end{center}

To begin with, DFT and the Kohn-Sham equations \cite{Martin} have been derived rigorously from the $N$-particle
Schr\"odinger equation of a system of interacting electrons moving in the external 
potential of the nuclei. For practical computations, an approximation has to be applied 
to the exchange-correlation energy functional $E_{\rm XC}[n]$. As correctly pointed out 
by SEA, the development of improved functionals $E_{\rm XC}[n]$ is still a matter of active current 
research. The role of $E_{\rm XC}[n]$ is to account for the exchange-correlation effects 
of the interacting electrons. All the single-particle quantum mechanical effects, however, 
are fully accounted for by solving the Kohn-Sham equations, which are formally equivalent 
to a single-particle Schr\"odinger equation for particles moving in an effective potential 
$v_{\rm eff}([n]; r)$ that is determined self-sonsistently. Thus, as denoted in Table 1 below, 
DFT correctly captures all quantum effects. It becomes accurate at high densities, i.e. in 
the weak coupling limit, when the mean interparticle distance ${\bar r}$ is much less than the Bohr radius $a_B$ (the scale of the local field)~\cite{density}, i.e. 
the Brueckner parameter $r_s={\bar r}/a_B$ is much less than one. At lower densities its
accuracy is determined by the exchange-correlation functional $E_{\rm XC}$. For hydrogen 
reliable expressions for $E_{\rm XC}[n]$ exist which were discussed in Ref. \cite{bonitz_comment} allowing 
for accurate calculations of the proton potential in a dense jellium background. In 
particular, bound states between protons and Friedel oscillations are correctly reproduced.
%\normaltext

Compared to DFT, QHD and LQHD contain three important additional approximations: 
\begin{description}
 \item[({\it i})] instead of a set of in general complex wave functions, QHD and LQHD solve for 
 real-valued hydrodynamic quantities, thereby, losing access to quantum interference 
 effects (2), in contrast to the statements of SEA \cite{shukla_comment};
 \item[({\it ii})] as any hydrodynamic theory, QHD averages over a finite volume containing many  
 particles, thereby losing the capability to resolve scales on the order of the mean 
 inter-particle distance ${\bar r}$, cf. point (7), the relevant length scale was determined 
 in Ref. \cite{bonitz_comment};
 \item[({\it iii})] SEA correctly note that QHD is a free electron theory, i.e. it does not contain electron-electron
 correlation effects and can only be applied to high densities, $r_s \lesssim 1$ (indicated 
 by the shaded area in Fig. 1 of Ref.~\cite{bonitz_comment}), cf. point (6) in the table. 
 Surprisingly, in their Letter \cite{shukla_11} Shukla and Eliasson extended their claim of 
 the attractive potential to the low-density value of $r_s \sim 26$ supporting it by numerical data. Comparing their data 
to DFT simulations we demonstrated \cite{bonitz_comment} that no ``novel'' potential minimum other 
than caused by bound states or Friedel oscillations exists.
 \end{description}

In conclusion, even though DFT is -- obviously -- not an exact theory, as correctly pointed 
out by SEA, it is nonetheless generally more accurate than QHD (with or without linearization) by 
construction. Therefore, a ``failure of DFT'' \cite{shukla_comment} to reproduce predictions 
of linearized quantum hydrodynamics (LQHD) is a serious indication of the failure of the latter.

This work was supported by the Deutsche Forschungsgemeinschaft via SFB-TR 24 project A5.

% and a grant for computing time at the North-German Supercomputing Alliance (HLRN). 


\begin{thebibliography}{33}
\bibitem{density} As in Ref.~\cite{bonitz_comment} the plasma density $n$ is characterized by the common Brueckner parameter, $r_s= {\bar r}/a_B$, where $a_B$ denotes the Bohr radius and ${\bar r}$ the mean interparticle distance which is related to the density by $4\pi {\bar r}^3/3 = n^{-1}$.
\end{thebibliography}
\end{document}